# GalliformeSpectra: A Hen Breed Dataset

Galib Muhammad Shahriar Himel[*a,b,c], Md Masudul Islam[d,e]

[a] School of Computer Sciences, Universiti Sains Malaysia, George Town, Penang, Malaysia
[b] Department of Computer Science, American International University-Bangladesh, Dhaka, Bangladesh
[c] Department of Computer Science, University of Greenwich, London, United Kingdom
[d] Department of Computer Science and Engineering, Bangladesh University of Business and Technology (BUBT), Dhaka, Bangladesh
[e] Department of Computer Science and Engineering, Jahangirnagar University, Dhaka, Bangladesh

*galib.himel.phd@student.usm.my
*galib.muhammad.shahriar@gmail.com

**Abstract**
This article presents a comprehensive dataset featuring ten distinct hen breeds, sourced from various regions, capturing the unique characteristics and traits of each breed. The dataset encompasses Bielefeld, Blackorpington, Brahma, Buckeye, Fayoumi, Leghorn, Newhampshire, Plymouthrock, Sussex, and Turken breeds, offering a diverse representation of poultry commonly bred worldwide. A total of 1010 original JPG images were meticulously collected, showcasing the physical attributes, feather patterns, and distinctive features of each hen breed. These images were subsequently standardized, resized, and converted to PNG format for consistency within the dataset. The compilation, although unevenly distributed across the breeds, provides a rich resource, serving as a foundation for research and applications in poultry science, genetics, and agricultural studies. This dataset holds significant potential to contribute to various fields by enabling the exploration and analysis of unique characteristics and genetic traits across different hen breeds, thereby supporting advancements in poultry breeding, farming, and genetic research.

## SPECIFICATIONS TABLE

| Subject | Computer Science |
|---|---|
| Specific subject area | Computer Vision, Pattern Recognition, machine learning, deep learning |
| Data format | Raw |
| Type of data | JPG, PNG Image |
| Data collection | To create this dataset, images were captured from various farms located in Germany, the UK, the USA, Egypt, Italy, and Romania and some images of the respected breeds were captured from Bangladesh. Many professionals around the world have captured these images and stored them in public cloud databases. Then we collected those data from the cloud repository and organized them breed-wise. In some cases, we have cropped some portions of the images for unnecessary background removal and also applied some noise reduction while necessary. |
| Data source location | **Location:** Bangladesh, Germany, UK, USA, Egypt, Italy, and Romania. |
| Data accessibility | **Repository 1 name:** Mendeley Data<br>**Data identification number:** 10.17632/nk3zbvd5h8.1<br>**Direct URL to data**: https://data.mendeley.com/datasets/nk3zbvd5h8/1<br>**Repository 2 name:** Science Data Bank<br>**Data identification number:** 10.57760/sciencedb.12798<br>**Direct URL to data**: https://www.scidb.cn/en/s/aA7BVf |

# VALUE OF THE DATA

- This comprehensive hen breed dataset holds immense value for the development of machine learning models, specifically in the domain of animal classification and recognition. It serves as a foundational resource for creating automated systems capable of accurately identifying and categorizing diverse hen breeds.
- Agricultural researchers and poultry scientists can leverage this dataset to advance the understanding of genetic traits, physical characteristics, and behavioral patterns specific to each hen breed. It aids in the identification and differentiation of various poultry types, contributing significantly to studies in animal husbandry, breeding, and genetics.
- The dataset's applicability extends to farm management and disease detection within poultry farming. Farmers and veterinarians can utilize this resource to improve disease diagnosis, implement effective breeding programs, and optimize flock management strategies based on the distinct traits and susceptibilities of each hen breed.
- In the realm of computer vision and image recognition, this dataset acts as a valuable resource for training and validating models aimed at automating processes in poultry farming. By accurately identifying and categorizing hen breeds, it assists in streamlining tasks such as sorting, breeding, and monitoring within the industry.
- The dataset's diversity of hen breeds enables the development of sophisticated classification models, aiding farmers in making informed decisions regarding breed selection, resource allocation, and breeding practices. It presents an opportunity for creating predictive models that optimize breeding programs and enhance overall productivity within the poultry sector.
- Researchers and scientists, particularly in the fields of animal science, genetics, and agriculture, can utilize this dataset to explore and understand the unique characteristics, genetic variations, and behavioral attributes of diverse hen breeds. It offers a platform for further studies and innovations in poultry research and management.

# DATA DESCRIPTION

Hen breeds represent a rich tapestry of avian diversity, each breed distinguished by unique traits, plumage, and characteristics. This diversity holds significance not only in agriculture but also in cultural heritage and genetic preservation. The preservation and proper management of these breeds are essential for maintaining genetic diversity and ensuring the resilience of poultry populations. Challenges such as disease management, conservation, and sustainable breeding practices are critical issues in the poultry industry. Moreover, recognizing and preserving these breeds plays a vital role in the conservation of genetic resources, promoting biodiversity, and addressing food security concerns. The Hen Breed Dataset includes ten distinct poultry breeds, such as Bielefeld, Blackorpington, Brahma, Buckeye, Fayoumi, Leghorn, Newhampshire, Plymouthrock, Sussex, and Turken. Each breed is depicted in a collection of images, showcasing their individual characteristics and diverse origins. This dataset is a valuable resource for accurate breed classification and serves as a visual reference for deep learning models. It comprises a total of 5050 images, encompassing original, resized, and augmented versions. Augmented images are crucial for effective model development, as machine vision relies on extensive training data. The dataset contributes to advancements in poultry research and breed recognition applications.

In our research paper, we introduced three variations of our dataset: the Original Dataset, the Resized Dataset, and the Resized Augmented Dataset. Each variation is further subdivided into ten sub-folders representing the specific breeds. The Original Dataset contains a total of 1,010 raw JPG images, with varying dimensions and a file size of 115 MB. The Resized Dataset, converted to PNG images, maintains a uniform resolution of 224 x 224 pixels, resulting in an initial file size of 97.5 MB. The Resized

Augmented Dataset, featuring 5050 PNG images, also consistently formatted at 224 x 224 pixels, results in a file size of 465 MB.

Our dataset's folder structure is visually depicted in **Fig. 1**, while the dataset creation process is illustrated in **Fig. 2**. This dataset is conveniently accessible through Mendeley Data [1] and Science Data Bank [2], stored in three separate zip files: 'Original_Dataset.zip,' 'Resized_Augmented_Dataset.zip,' and 'Resized_Dataset.zip.'

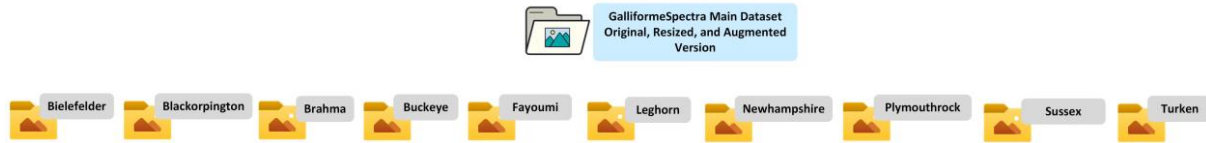

**Fig. 1.** Dataset Folder Structure

Here are some identification features [3] for each of the hen breeds:

1. **Bielefeld**: Bielefeld hens are known for their striking appearance with black and white plumage. They have a medium-sized, well-rounded body, and their face and comb are red.
2. **Blackorpington**: Black Orpingtons are large, dual-purpose birds with glossy black feathers. They have a broad, round body, and their skin and legs are pinkish-white. Their combs are red and their eyes are dark.
3. **Brahma**: Brahmas are large, heavy birds with feathered shanks and toes. They have a striking appearance with their black and white plumage and a pea comb. Their eyes are bay and their beaks are yellow.
4. **Buckeye**: Buckeye hens are known for their mahogany-red feathers, which are almost uniform in color. They have a muscular body, red wattles, and a pea comb.
5. **Fayoumi**: Fayoumi hens are small and active birds with distinct black and white mottled plumage. They have a rose comb, red earlobes, and a lean body.
6. **Leghorn**: Leghorns are slender, small to medium-sized birds known for their white plumage. They have single combs, red earlobes, and bright red wattles.
7. **Newhampshire**: New Hampshire hens are deep red in color with a single comb. They have a well-rounded body and are known for their egg-laying capabilities.
8. **Plymouthrock**: Plymouth Rock hens have black and white striped plumage and a single comb. They are medium-sized birds with a friendly and docile temperament.
9. **Sussex**: Sussex hens have white feathers with black specks, and they have a single comb. They are medium-sized birds with a graceful appearance.
10. **Turken**: Turkens are unique with their partially featherless necks. They have red combs and wattles, and their feathers can come in various colors.

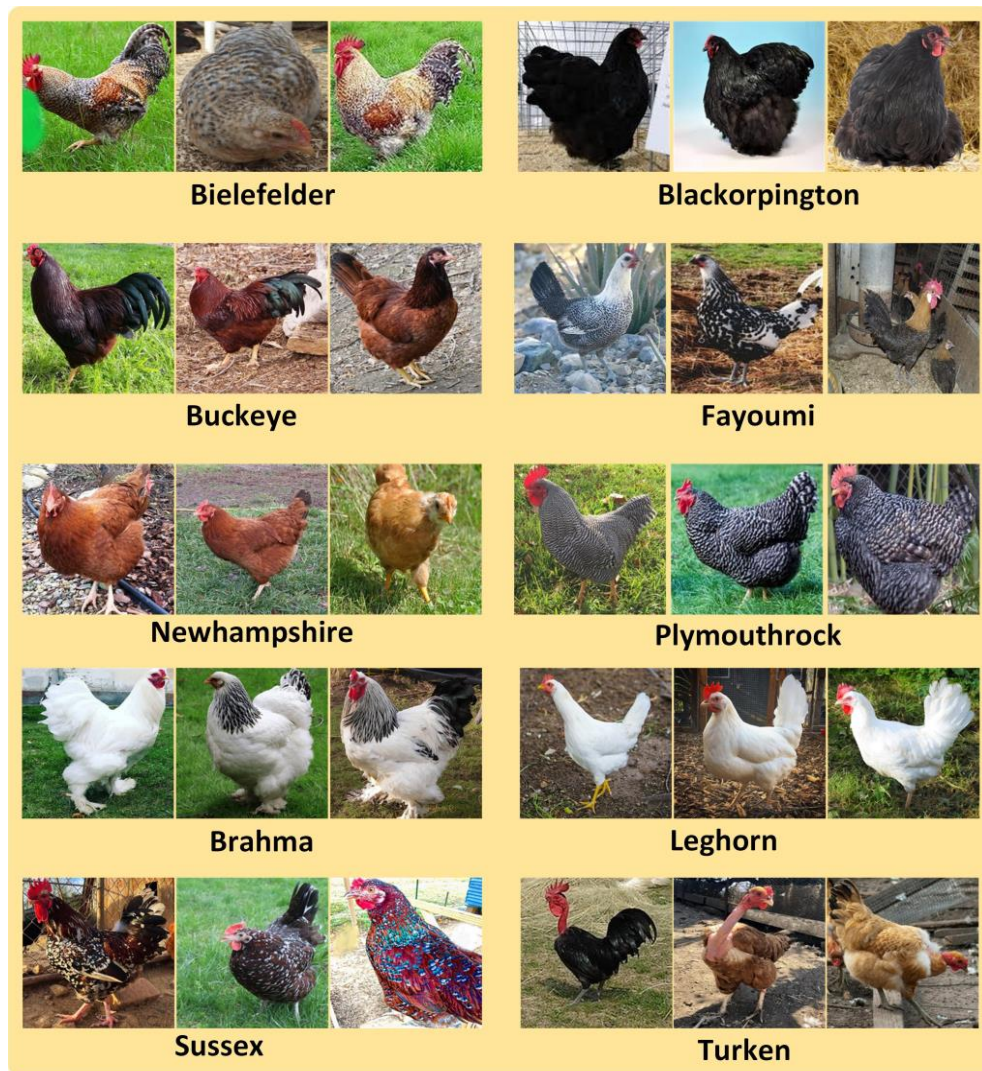

**Fig. 2.** Examples of Dataset

## EXPERIMENTAL DESIGN, MATERIALS AND METHODS

The process of gathering images for each distinct hen breed followed a structured methodology, as depicted in **Fig. 3**. The selection process for individual hens within each breed category was carried out with a focus on maintaining an even distribution, ensuring equal representation of all hen breeds within the dataset. This method aimed to capture a diverse array of visual attributes specific to each breed. To achieve this diversity, a systematic random selection approach was employed. Each hen selected for photography was chosen randomly from a cluster of hens representing a particular breed. Following the image capture, the photographs were systematically transferred from the capturing device, such as a camera or smartphone, to an external storage device for safekeeping. These images were then accurately organized into respective folders, each folder labeled with the name of the corresponding hen breed. After transferring and organizing the images, the process of capturing images for the next hen breed began once the images of the previously captured hen breed were securely stored and removed from the capturing device, ensuring a methodical and structured approach to dataset collection.

This systematic methodology continued until images for all ten categories of hen breeds were collected, encompassing various characteristics and stages of development, thereby ensuring a comprehensive representation of the dataset.

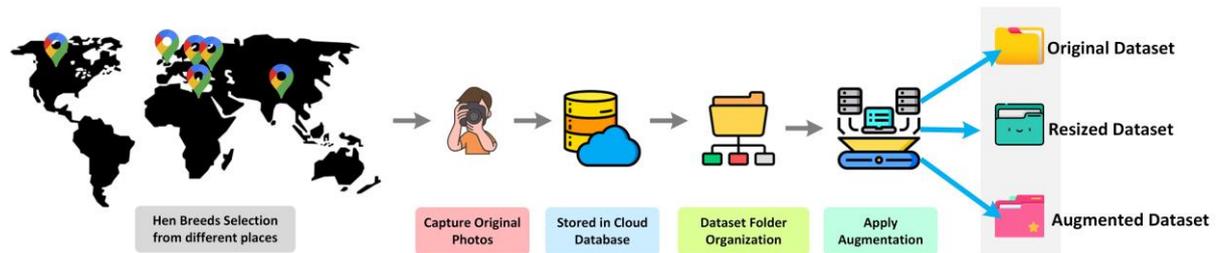

**Fig. 3.** Workflow for the Hen Breed Dataset Collection

Statistical insights regarding the distribution and quantity of images for each hen breed are presented in **Table 1**, providing a quantitative view of the dataset's composition and balance across the diverse hen breeds.

**Table 1.** Statistics of the Original & Augmented Hen Breeds Dataset

| Hen Breeds | Original Images | Augmented Images |
|---|---|---|
| Bielefeld | 101 | 505 |
| Blackorpington | 101 | 505 |
| Brahma | 101 | 505 |
| Buckeye | 101 | 505 |
| Fayoumi | 101 | 505 |
| Leghorn | 101 | 505 |
| Newhampshire | 101 | 505 |
| Plymouthrock | 101 | 505 |
| Sussex | 101 | 505 |
| Turken | 101 | 505 |
| **Total** | **1010** | **5050** |

# ETHICS STATEMENT

This article does not involve any research involving human or animal subjects by any of the authors. The datasets utilized in this article are publicly accessible. When utilizing these datasets, it is essential to adhere to appropriate citation guidelines.

# CRediT AUTHOR STATEMENT

**Galib Muhammad Shahriar Himel**: Investigation, Software, Validation, Formal Analysis, Resources, Data Curation, Visualization, Project administration, Supervision, Conceptualization, Methodology, Writing - Original Draft, Writing - Review and editing. **Md. Masudul Islam**: Data Curation, Project administration, Conceptualization, Methodology, Formal analysis, Resources, Writing - Original Draft, Visualization.

# DATA AVAILABILITY

[GalliformeSpectra: A Hen Breed Dataset](#) (Mendeley Data)
[GalliformeSpectra: A Hen Breed Dataset](#) (Science Data Bank)

## ACKNOWLEDGEMENTS

This research did not receive any specific grant from funding agencies in the public, commercial, or not-for-profit sectors.

## DECLARATION OF COMPETING INTERESTS

The authors declare that they have no known competing financial interests or personal relationships that could have appeared to influence the work reported in this paper.